\documentclass[11pt]{article}

\usepackage[letterpaper,margin=1in]{geometry}
\usepackage[T1]{fontenc}
\usepackage{amsmath,amssymb}
\usepackage{newtxtext,newtxmath}
\usepackage{upgreek}
\usepackage{graphicx}
\usepackage{natbib}
\usepackage[colorlinks=true,linkcolor=blue,citecolor=blue,urlcolor=blue]{hyperref}

\newcommand{\Rey}{\mbox{\textit{Re}}}
\IfFileExists{t1phv.fd}
  {\DeclareMathAlphabet{\mathsfbi}{T1}{phv}{b}{it}}
  {\DeclareMathAlphabet{\mathsfbi}{OT1}{cmss}{m}{sl}}
\newcommand{\slsQ}{\mathsfbi{Q}}

\makeatletter
\newcommand{\aff}[1]{\textsuperscript{#1}}
\newcommand{\affinst}[1]{\par\vspace{3pt}\textsuperscript{#1}\ignorespaces}
\newcommand{\affiliation}[1]{\gdef\@affiliation{\let\aff\affinst #1\par}}
\gdef\@affiliation{}
\newcommand{\corresau}[1]{\gdef\@corresau{\textbf{Corresponding author: }#1}}
\gdef\@corresau{}
\newcommand{\email}[1]{\href{mailto:#1}{#1}}

\renewcommand{\maketitle}{%
  \begin{center}
  {\LARGE\bfseries \@title \par}
  \vspace{1.2em}
  {\large \def\and{and }\@author \par}
  \vspace{0.6em}
  {\normalsize \@affiliation}
  \ifx\@corresau\@empty\else\vspace{0.4em}{\small \@corresau \par}\fi
  \end{center}
  \vspace{1.5em}
}
\makeatother

\title{Wake interactions drive synchronized vortex merging in a hovering quadcopter}

\author{Elias S. Pratschke\aff{1}, Claus C. Wolf\aff{2}, Daniel Schanz\aff{2}, Andreas Schröder\aff{2,3} \and Oliver T. Schmidt\aff{1}}

\affiliation{\aff{1}Department of Mechanical and Aerospace Engineering, University of California San Diego, La Jolla, CA 92093, USA
\aff{2}German Aerospace Center (DLR), Göttingen, Germany
\aff{3}Institute of Transport Technology, Brandenburg University of Technology, Cottbus, Germany}

\corresau{Oliver T. Schmidt, \email{oschmidt@ucsd.edu}}

\begin{document}
\maketitle

\begin{abstract}



The most energetic coherent structure of a hovering full-scale quadcopter is associated with a self-organizing process in which the individual rotor vortices synchronize their frequencies while undergoing merging events, yielding a globally correlated structure. We identify and characterize this phenomenon by applying spectral modal and conditional analyses to assimilated three-dimensional velocity data acquired via Shake-The-Box Lagrangian particle tracking \citep{wolf_volumetric_2024}. The dataset captures a high-$\Rey$, turbulent flow further complicated by time-varying rotor speeds stemming from active flight control, low-frequency vehicle drift, finite spatio-temporal resolution, and measurement uncertainty. Most coherent structures recover established single-rotor features such as tip vortices and their subharmonic pairing. The globally synchronized vortex merging manifests as a spectral peak at an incommensurate frequency below the rotor band, which cannot be explained by single-rotor aerodynamics, subharmonic instabilities, or band-to-band triadic interactions. Instead, conditional averaging provides evidence of the aforementioned intermittent, distinctly non-subharmonic vortex-merging process involving all four rotor wakes. Establishing whether or not this phenomenon is observed across different flight conditions and configurations remains speculative; however, the consistent characterization of the globally synchronized vortex merging using complementary frequency- and time-domain analyses despite experimental complexities, in particular rotor speed variations, demonstrates its robustness.


\end{abstract}

\section{Introduction}
\label{sec:headings}

Quadcopter configurations are amongst the most popular for a variety of autonomous flight applications. The operation of multiple rotors in close proximity and interactions with fuselage elements typically result in complicated, highly three-dimensional flow dynamics that often differ significantly from simpler, canonical flows around isolated components \citep{shukla_multirotor_2018, kostek_experimental_2024, kiran_structure_2026}. Nearly all stages of the design process can benefit from detailed insights into the aircraft wake, with aerodynamic performance and reduced noise emission being typical design objectives. 

Rotors generate flow patterns dominated by high vorticity, characterized primarily by wingtip and root vortices \citep{joukowsky_vortex_1912}. This vortex system is generally unstable and susceptible to linear and nonlinear processes, such as mutual inductance and elliptical instabilities, among others \citep{leweke_dynamics_2016, leweke_long-_2014, sarmast_mutual_2014, quaranta_long-wave_2015, quaranta_local_2019}. Depending on flow parameters such as the Reynolds number $\Rey$ and design factors such as geometry and loading, a combination of these instabilities typically deforms the wake, eventually breaking it down into turbulence. In multirotor aircraft, these dynamics are further enriched by interactions between the vortex systems of individual rotors \citep{zhou_experimental_2017}. 

Resolving these multi-rotor interactions requires high-fidelity, volumetric data. While scale-resolving simulations of such complex, high-$\Rey$ flows often incur large computational costs or rely on heavily calibrated turbulence models \citep{stanly_direct_2025, hosseini_direct_2016, dorange_high_2024, ventura_diaz_high-fidelity_2018}, experimental data acquisition faces its own hurdles. Because physical probe arrays heavily disturb the flow field \citep{citriniti_reconstruction_2000}, non-intrusive imaging is essential. To this end, numerous particle image velocimetry (PIV, \cite{raffel_particle_2018}) and particle tracking velocimetry (PTV, \cite{nishino_three-dimensional_1989}) approaches have been developed, with Lagrangian Particle tracking (LPT) having recently gained popularity \citep{schroder_3d_2023}. The first combined usage of the Shake-The-Box (STB, \citep{schanz_shake--box_2016}) algorithm with two-stage scanning to produce time-resolved flow fields of the turbulent wake of a full-scale quadcopter is presented in \citep{schanz_scanning_2024,wolf_volumetric_2024}.

Although such well-resolved flow field data contain all dynamic quantities of interest, researchers and engineers are usually interested in processes such as structural vibrations, lift fluctuations, or aeroacoustics. Coherent structures often play an important role in such processes. Modal decompositions aim to combine the computation of coherent structures with physical interpretability by attaching an energy measure or other means of selection to the computed structures \citep{taira_modal_2017}. Well-known techniques include Proper Orthogonal Decomposition (POD, \cite{lumley_stochastic_1970,berkooz_proper_1993}) and Dynamic Mode Decomposition (DMD, \cite{schmid_dynamic_2010, rowley_spectral_2009}). The POD provides optimal basis functions ranked in terms of an energy norm, although the modes are generally not coherent in time. In contrast, the DMD algorithm attaches complex frequencies to the computed modes, fully describing their linear dynamics both in space and time. Both methods can be used for physical discovery and for the synthesis of reduced order models (ROM) \citep{rowley_model_2017}. 
Spectral POD (SPOD) \citep{towne_spectral_2018} provides a method that combines the main advantages of both POD and DMD for statistically stationary data. The computed modes are coherent in space and time, and monochromatic, allowing for an efficient separation of time scales. At each frequency, the SPOD provides a set of spatially orthogonal modes ranked in terms of an energy norm. A further extension to harmonically forced systems is presented in \citep{heidt_spectral_2024}.

Applications of modal analysis techniques to rotor-driven flows include works on wind turbine \citep{cherubini_data_2021,debnath_towards_2017,sarmast_mutual_2014} and marine propeller wakes \citep{magionesi_modal_2018}. Most works rely on the established POD and DMD algorithms. DMD is particularly effective at resolving the typical tonal dynamics of rotor flows that generally contain frequencies dictated by shaft speed and blade count. Applications to full-scale rotorcraft are sparse, due to the aforementioned difficulties associated with obtaining a well-resolved dataset.


This work applies modal decomposition techniques to a large-scale dataset of a quadcopter in hover flight \citep{wolf_volumetric_2024}. We use SPOD to identify dominant coherent structures and their frequencies, while a frequency-time approach \citep{nekkanti_frequencytime_2021} captures transient flow phenomena. Our results reveal two distinct vortex merging processes: a subharmonic tip-vortex pairing that connects the blade-passing and rotor frequencies of individual rotors, as well as an irregular, non-subharmonic process involving the wakes of all four rotors. The latter collective interaction occurs at an incommensurate frequency, $f_{\star}$ which emerges as the most energetic spectral peak apart from the zeroth frequency bin. We demonstrate that this event represents a synchronization process driven by the mutual coupling of the individual wakes, reminiscent of phenomena documented in both fundamental dynamical models \citep{rand_bifurcation_1980} and nonlinear fluid flows \citep{provansal_benard-von_1987, bonciolini_low_2021, herrmann_modeling_2020}.

\section{Data and Methods}

\subsection{Dataset}
\label{sec:dataset}

Experimental flow field data of a commercial DJI Mavic quadcopter hovering out of ground effect were acquired using Shake-The-Box LPT. To obtain smooth high-resolution flow fields, approx.~400\,000 instantaneously tracked particles were processed using the FlowFit 3 data assimilation routine \citep{gesemann_noisy_2016, godbersen_flowfit3_2024}, which enforces incompressible Navier--Stokes constraints. FlowFit 3 performs a regularized fit of the particle velocities and accelerations to a staggered grid from cubic B-splines. The latter spatially continuous functions can be evaluated at any point within the measurement volume and thus sampled at arbitrary locations.

Full details of both the experiment and LPT methodology are provided in \cite{wolf_volumetric_2024} and \cite{schanz_scanning_2024}, respectively. Particle tracking was performed at a recording frequency of 3000 Hz. FlowFit was applied at every fourth time-step, resulting in a sampling rate of 750 Hz ($\upDelta t \approx 0.0013$ seconds). The total time series, comprising 1964 snapshots, spans 2.62 seconds. The flow fields were sampled on a uniform Cartesian grid of 136×146×220 points. 

\begin{figure}
  \centerline{\includegraphics{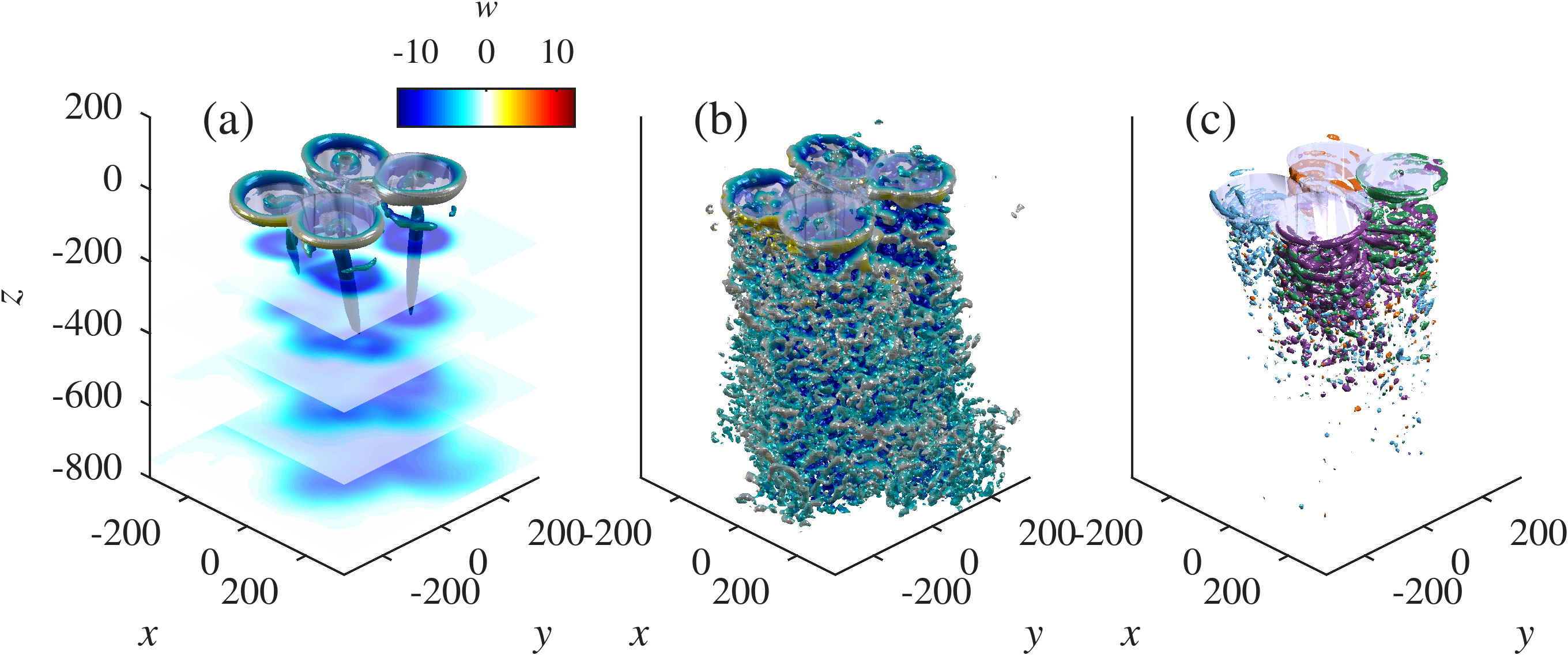}}
  \caption{\textit{(a)}: Q-criterion isocontour of the mean flow field, along with slices of the streamwise velocity $w$. \textit{(b)}: Q-criterion isocontour of the flow field at $t = 283 \upDelta t$. On both contours, the colors represent $w$. \textit{(c)}: sparse reconstruction of the flow field at $t = 283\upDelta t$. Flow fields of individual rotors are approximately separated and colored. The same color coding is used throughout the work.}
\label{fig:snapshot_mean}
\end{figure}

Figure \ref{fig:snapshot_mean}(a) visualizes the mean flow, where the Q-criterion isosurface shows both an inner and outer vortex ring alongside vortex cores extending downstream from each rotor. Slices of the streamwise velocity at four axial locations illustrate the transition of the mean flow from high-gradient features near the rotor disks to a more diffuse wake further downstream. An instantaneous snapshot of the flow field is provided in figure \ref{fig:snapshot_mean}(b), with the Q-criterion isosurface (colored by streamwise velocity) highlighting the rich, chaotic vortex dynamics within the wake. Finally, figure \ref{fig:snapshot_mean}(c) displays a sparse reconstruction of the flow using the leading SPOD modes. By isolating frequencies associated with specific rotors, the flow field is decomposed into individual rotor contributions and color-coded accordingly. The color scheme of orange I, green II, blue III, and purple IV for the different rotors is maintained throughout the rest of this work.

Multiple disturbance factors led to a slow, random drift of the quadcopter during the data acquisition process. The average drift velocity was $0.034$ m/s, with the highest value being $0.105$ m/s, which is on the order of a percent of the induced velocities of the rotors. A large portion of this drift was removed during post-processing, but a small amount of relatively slow, random motion remained in the dataset. Finally, the tracked particle density close to the rotor planes was comparatively low, due to the rotors and support beams blocking the line of sight of the cameras. The velocity values in those regions of the flow field thus rely mostly on the data assimilation routine.

Both rotor and blade-passing frequencies (BPF) often emerge as ubiquitous spectral features in frequency-domain analyses of rotor wakes. For the two-bladed configuration used in this work, the BPF are exactly twice the rotor frequencies. Based on the drone telemetry, the average rotor frequencies were 94 Hz for the front rotors, denoted II and IV, and 81 Hz for the rear rotors, denoted I and III. While the drone was nominally in hover flight, the flight controller performed subtle, continuous adjustments of the rotor speeds to maintain stability. We analyze how these dynamic corrections influence the observed frequency characteristics in Section~\ref{sec:frequency-time}.

\subsection{Spectral Proper Orthogonal Decomposition}
\label{sec:spod}

In this section, we provide a brief introduction to the SPOD and the related methods used to generate the results in this work. A comprehensive derivation of the SPOD, including its relationship to DMD and resolvent analysis, can be found in \cite{towne_spectral_2018}, while best practices for parameter selection are outlined by \cite{schmidt_guide_2020}. SPOD seeks modes that optimally capture space-time flow dynamics in a statistically stationary, ergodic flow. These SPOD modes are obtained as the eigenfunctions of the cross-spectral density (CSD) matrix $\mathsfbi{S}(x,x^{\prime},f) = \int_{-\infty}^{\infty} \mathsfbi{C}(x,x^{\prime},\tau)e^{-2\text{i}\pi f\tau} \text{d}\tau$, which in turn is the Fourier transform of the space-time correlation tensor $\mathsfbi{C}(x,x^{\prime},\tau)$, where $\tau = t - t^{\prime}$. The eigenvalue problem is stated as follows, both in continuous (equation \ref{eq:spod_cont}) and discrete (equation \ref{eq:spod_disc}) form:

\begin{eqnarray}  
    \int_{\Omega} \mathsfbi{S}(x,x^{\prime},f) \mathsfbi{W}(x^{\prime}) \boldsymbol{\phi}(x^{\prime},f) \text{d}x^{\prime} = \boldsymbol{\phi}(x,f) {\lambda}(f) \label{eq:spod_cont} \\
        \frac{1}{N_{\text{blk}}}\hat{\slsQ}^{H}_{j} \mathsfbi{W} \hat{\slsQ}_{j} {\boldsymbol{\Psi}}_{j} = \boldsymbol{{\Psi}}_{j}\boldsymbol{\Lambda}_{j}, \hspace{0.1in} \boldsymbol{{\Phi}}_{j} = \frac{1}{\sqrt{N_{\text{blk}}}}\hat{\slsQ}_{j}\boldsymbol{{\Psi}}_{j} \boldsymbol{\Lambda}_{j}^{-\frac{1}{2}}. \label{eq:spod_disc}
\end{eqnarray}
For the discrete problem, the CSD is estimated at each frequency using Welch's method \citep{welch_use_1967}, by segmenting the dataset into $N_{\text{blk}}$ overlapping blocks, which are then treated as separate realizations. Averaging the fast Fourier transform (FFT) of these blocks yields a sample of $\mathsfbi{S}(x,x^{\prime},j \cdot \upDelta f) \approx \hat{\slsQ}^{H}_{j} \mathsfbi{W} \hat{\slsQ}_{j}$ at each discrete frequency bin $j$.

The $k$ eigenmodes obtained from equation \ref{eq:spod_disc}, denoted $\phi_{j}^{k}, k = 1,...,N_{\text{blk}}$ at frequency $j$ are ranked by their eigenvalues $\lambda_{k}$, representing the energy in the norm induced by the weighted inner product $\langle \boldsymbol{q}_{i}, \boldsymbol{q}_{j} \rangle = \boldsymbol{q}_{i}^{\text{T}} \mathsfbi{W} \boldsymbol{q}_{j}$. The positive semi-
definite weight matrix $\mathsfbi{W}$ can also be used to emphasize specific regions of the computational domain. Notably, even if parts of the domain are assigned a weight of zero (masking), the resulting modes still represent optimally correlated structures throughout the entire flow field \citep{boree_extended_2003}. The SPOD modes can be used to reconstruct the original dataset, and low-order representations can be achieved by retaining only a subset of leading modes and frequencies \citep{nekkanti_frequencytime_2021}:

\begin{eqnarray}
    \mathsfbi{A}_j = \sqrt{N_{\text{blk}}} \boldsymbol{\Lambda}_{j}^{\frac{1}{2}} \boldsymbol{\Psi}^{H}_{j}, \hspace{0.2in} \hat{\slsQ}_j = \boldsymbol{\Phi}_{j}\mathsfbi{A}_{j} \\
    \slsQ = \mathcal{F}^{-1} \left( \left[ \hat{\slsQ_{1}}, \hat{\slsQ_{2}}, ..., \hat{\slsQ}_{N_{\text{DFT}}} \right]\right).
\end{eqnarray}
To extract continuous, time-varying expansion coefficients across all of the overlapping blocks, we project the data onto the SPOD basis, using the frequency-domain approach outlined in \cite{nekkanti_frequencytime_2021}. For this method, a short-time Fourier transform is used. The following expression is obtained in continuous time, for the $k$-th mode:

\begin{eqnarray}
    a^{(k)}(f,t) = \int_{\Omega}  \int_{\upDelta T} \boldsymbol{\phi}^{(k)} (x, f)^{H} \mathsfbi{W}(x) \boldsymbol{q}(x,t + \tau)w(\tau)e^{-2\text{i}\pi f \tau} \text{d}x\text{d}\tau.
\end{eqnarray}
In the case of discrete flow data, the above convolution is most efficiently computed in two steps, with the projection of the snapshots onto the modal basis being the first:

\begin{eqnarray}
    \tilde{a}_{j}^{k}(t_{i}) = \langle  \boldsymbol{q}(x,t_i)', \phi(x)_{j}^{k} \rangle,
\end{eqnarray}
where $\phi_{j}^k$ is the $k$-th basis function obtained at frequency bin $j$ via SPOD. Next, we perform the following convolution to obtain the temporal dynamics of the expansion coefficients, which can be efficiently achieved using the FFT \citep{chu_stochastic_2025}:

\begin{eqnarray}
    a_j^{k}(t_i) = \sum_{n=0}^{N_\text{DFT}-1} w(n) \cdot \tilde{a}_{j}^{k}(t_i + n) \cdot e^{-\text{i} 2\pi n\upDelta t j \upDelta f},
\end{eqnarray}
where the function $w$ accounts for any windowing used in the computation of the SPOD. This procedure yields smooth, complex-valued coefficients $a_j^{k}(t_i)$ associated with the monochromatic flow structures described by the SPOD modes. Using the magnitude of these coefficients, we can construct spectrograms that reveal the frequency-time evolution of the flow.

\section{Results}
\label{sec:results}
The dataset was partitioned into $N_{\text{blk}} = 6$ blocks, each containing $N_{\text{DFT}} = 512$ snapshots that overlapped by $50\%$. A Hamming window was used. All computations were performed using the three velocity components of the flow as the state vector.
\subsection{Modal frequency-time dynamics}
\label{sec:frequency-time}

\begin{figure}
  \centerline{\includegraphics{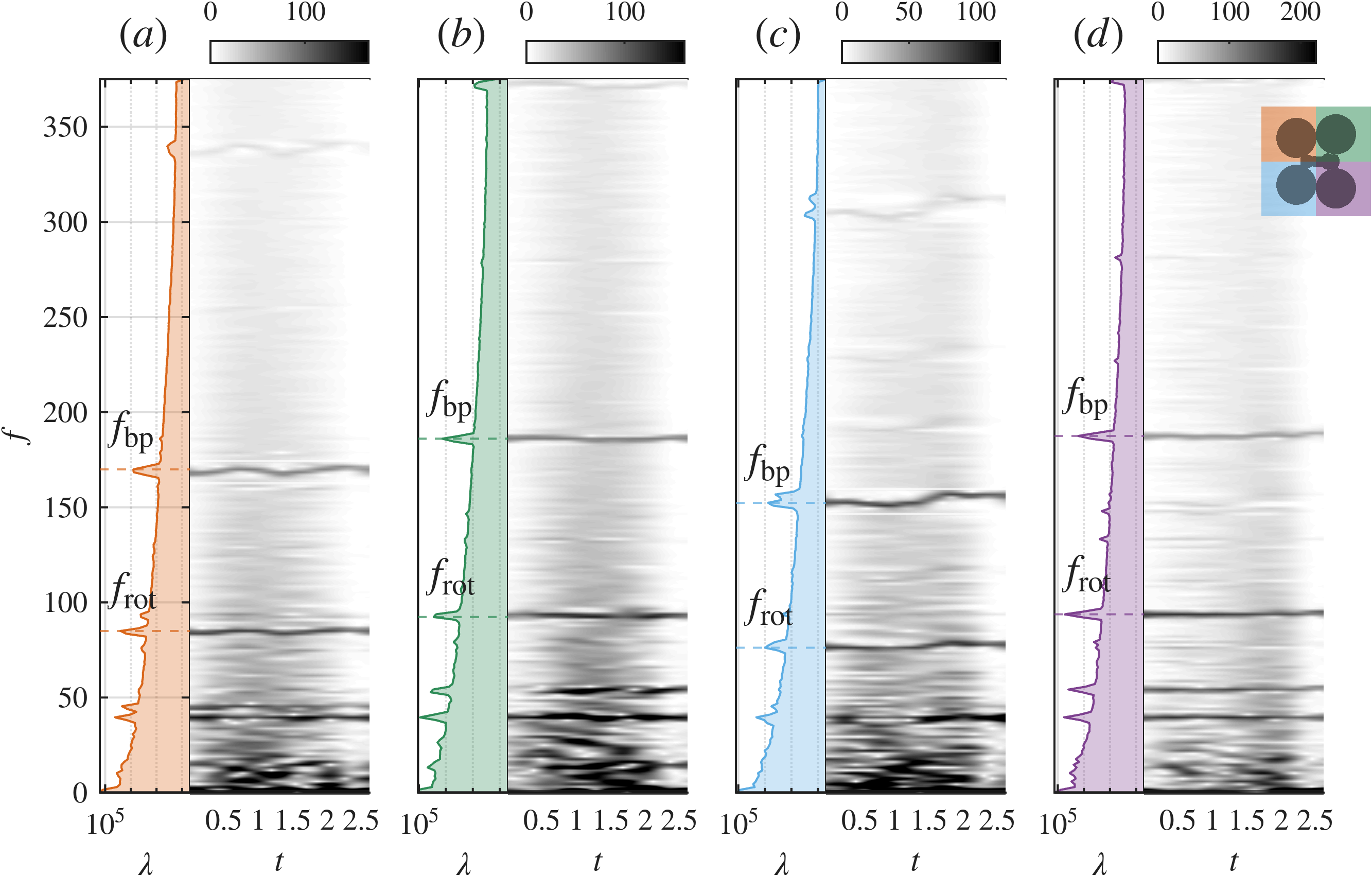}}
  \caption{Scalograms of the leading expansion coefficient $\vert a_{1}^{j}(t)\vert$ produced by applying frequency-time analysis to the SPOD modes of each individual rotor (\textit{(a)}): I, \textit{(b)}: II, \textit{(c)}: III, \textit{(d)}: IV). The insets to the left of each scalogram show the magnitude of the leading SPOD eigenvalue. The drone schematic in \textit{(d)} shows the colors chosen to represent the different rotors.}
\label{fig:scalograms}
\end{figure}

Because experimental measurements rarely satisfy the strict stationarity assumed by SPOD, we evaluate the time-dependency of our dataset using the frequency-time (FT) analysis method introduced in Section~\ref{sec:spod}. A high magnitude of the leading expansion coefficient indicates a strong similarity between the instantaneous flow field and the first SPOD mode at a given frequency. Individual rotor dynamics are isolated by applying four quadrant-wise weighting matrices, shown schematically in figure~\ref{fig:scalograms}\textit{(d)}.

The resulting scalograms in figure \ref{fig:scalograms}\textit{(a--d)} reveal different behaviors across the rotors. For Rotor~I (figure~\ref{fig:scalograms}\textit{(a)}), the blade-passing frequency $f_{\mathrm{bp}}^{\text{I}}$ exhibits a slow, periodic drift that accounts for the spectral broadening around $f_{\mathrm{bp}}^{\text{I}}$ in the SPOD eigenvalue spectrum. Similar variation occurs at $f_{\mathrm{rot}}^{\text{I}}$ and $2f_{\mathrm{bp}}^{\text{I}}$, though $2f_{\mathrm{bp}}^{\text{I}}$ exhibits a lower magnitude due to its lower mode energy. In contrast, Rotor~III (figure~\ref{fig:scalograms}\textit{(c)}) indicates a rapid shift to higher frequencies at $t \approx 1.5~\text{s}$ in both $f_{\mathrm{bp}}^{\text{III}}$ and $f_{\mathrm{rot}}^{\text{III}}$. The frequencies before and after this transition correspond to the two peaks around $f_{\mathrm{bp}}^{\text{III}}$ and the broad $f_{\mathrm{rot}}^{\text{III}}$ peak in the global SPOD spectrum. This shift persists at $2f_{\mathrm{bp}}^{\text{III}}$, though it is only barely discernible due to the low energy of the modes. Meanwhile, Rotors~II and IV (figure~\ref{fig:scalograms}\textit{(b, d)}) remain nearly stationary throughout the experiment.

Below the rotor frequencies, all four rotors display energetic structures characterized by varying degrees of transient behavior that lack a clear pattern. The consistently elevated magnitude in the zeroth frequency bin stems from low-frequency drifts below the window resolution, which manifest as spectral leakage into the zeroth bin.

These observations highlight the importance of interpreting the power spectrum and, by extension, the SPOD eigenvalue spectrum, from a purely statistical perspective. Although time-varying frequency content technically renders the flow non-stationary, frequency-domain methods remain valid provided these transient dynamics are explicitly accounted for. The clearest signature of this transience is the appearance of multiple eigenvalue peaks for a single physical phenomenon, as observed at the BPF of Rotor III. When frequency changes are too small to result in separate peaks, spectral broadening is observed instead. Because truncating the dataset to exclude Rotor III's frequency switch would still leave Rotor I's spectral broadening intact, we retain the full dataset and verify these transient phenomena a posteriori.

\subsection{Spectra \& Structures}
\label{sec:spectra_structures}

\begin{figure}
  \centerline{\includegraphics{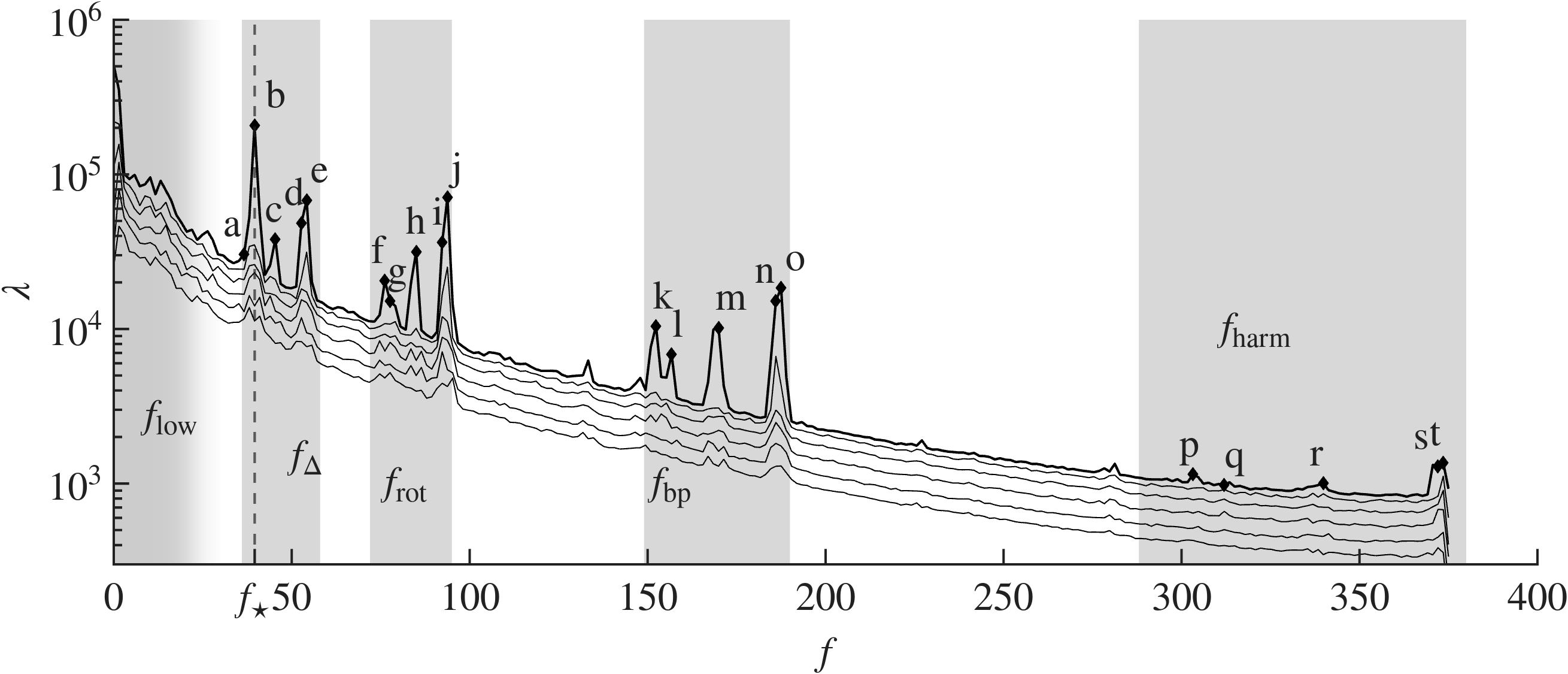}}
  \caption{SPOD eigenvalue spectrum computed using uniform weighting matrix. The modes corresponding to peaks marked with letters are shown in figure \ref{fig:combined_modes}.}
\label{fig:combined_spect}
\end{figure}

\begin{figure}
  \centerline{\includegraphics{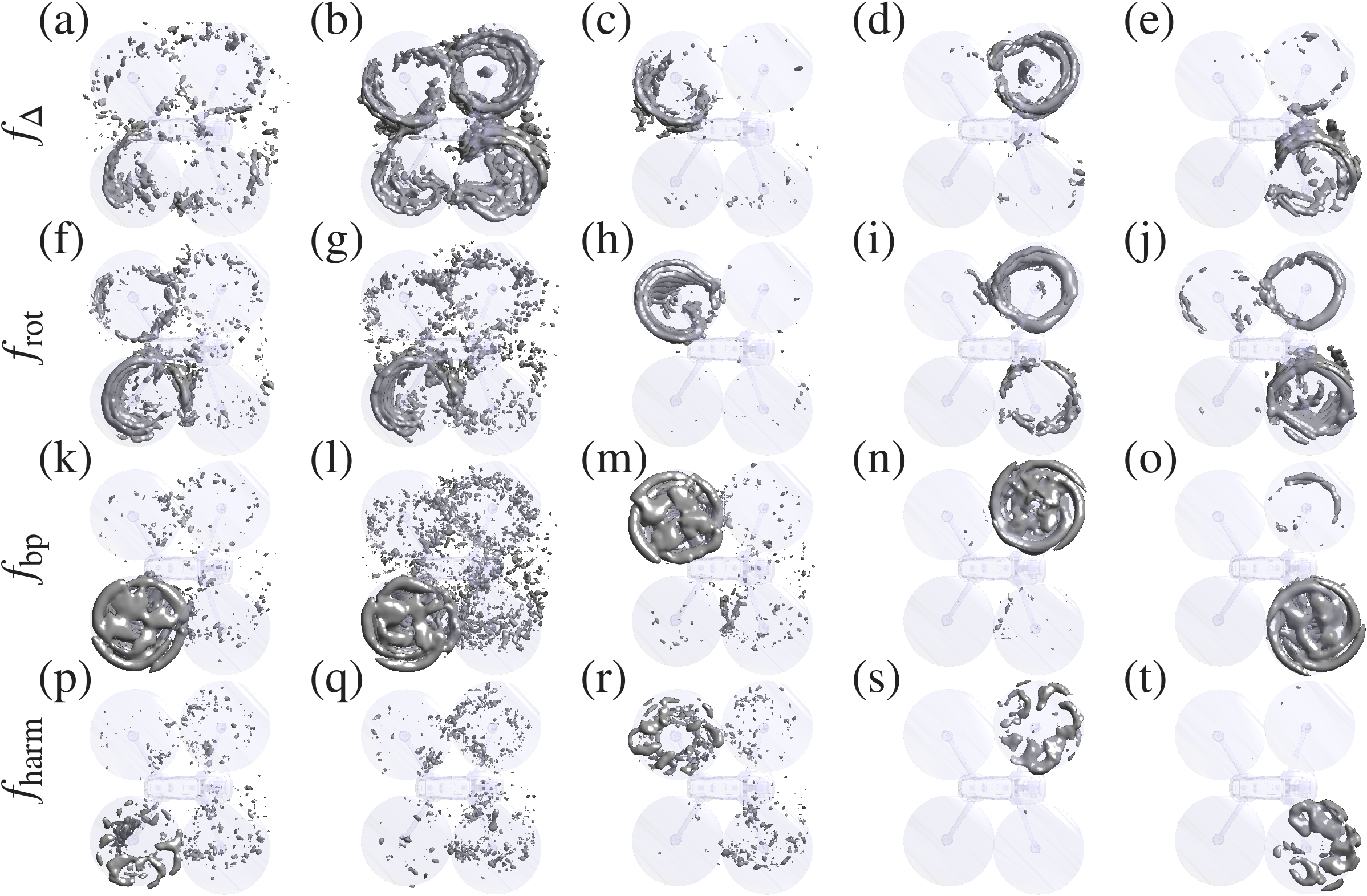}}
  \caption{Q-criterion isosurfaces of the modal structures associated with the marked peaks in the eigenvalue spectrum shown in figure \ref{fig:combined_spect}. The values for the isosurfaces were selected on a mode-by-mode basis as to increase interpretability.}
\label{fig:combined_modes}
\end{figure}

With the insights from the FT analysis in mind, we apply SPOD to the dataset using a uniform weighting matrix, including the entire flow field to an equal degree in the energy norm. The resulting eigenvalue spectrum is shown in figure \ref{fig:combined_spect}. The spectrum displays the magnitude of the $N_{\text{blk}}$ ranked eigenvalues, representing the energy of their corresponding modal structures at each frequency. Representative Q-criterion isosurfaces of the leading modes are displayed in figure \ref{fig:combined_modes}\textit{(a--t)}, with their corresponding eigenvalues identified by matching letters in figure \ref{fig:combined_spect}.

Fundamental aerodynamic principles govern the shedding of wingtip and root vortices \citep{joukowsky_vortex_1912}. For a rotor, this shedding occurs at the BPF, establishing it as a fundamental frequency of the system. The $f_{\text{bp}}$ modes (figure \ref{fig:combined_modes}\textit{(k--o)}) display clear spatial organization across individual rotors. The two lowest-frequency modes in this band (figure \ref{fig:combined_modes}\textit{(k, l)}) correspond to Rotor III; their closely spaced spectral peaks reflect the peak splitting noted in Section~\ref{sec:frequency-time}, while their low rank, characterized by a distinct spectral gap, confirms that these tonal structures are well-converged and separated from the broadband background. Similarly, the mode for Rotor~I (figure~\ref{fig:combined_modes}\textit{(m)}) is well-converged despite the spectral broadening. The remaining higher-frequency modes in this band (figure~\ref{fig:combined_modes}\textit{(n, o)}) are localized to Rotors II and IV, respectively.

The rotor shaft frequencies $f_{\text{rot}}$ likewise emerge as a set of prominent peaks, yielding modes characterized by well-defined helical vortices (figure \ref{fig:combined_modes}\textit{(f--j)}). As in the $f_{\text{bp}}$ band, the two lowest-frequency modes in this group (figure \ref{fig:combined_modes}\textit{(f, g)}) pertain to Rotor III; they are visually nearly identical and exhibit low rank, though their eigenvalues do not form separate peaks. Figure~\ref{fig:combined_modes}\textit{(h)} displays the corresponding localized structure for Rotor I, while figure \ref{fig:combined_modes}\textit{(i, j)} captures helical modes emphasizing Rotor II and Rotor IV at the lower and higher frequency, respectively.

The highest frequency band, $f_{\text{harm}} = 2f_{\text{bp}}$, contains the first BPF harmonics, with the harmonic frequency of Rotor IV coinciding with the Nyquist frequency. Although frequency-time analysis predicted two distinct harmonic frequency peaks for Rotor~III, these spectral features remain barely discernible, with minimal spectral gaps. Consequently, while Rotor~III’s pre-shift harmonic mode is moderately well-defined (figure~\ref{fig:combined_modes}\textit{(p)}), its post-shift counterpart (figure~\ref{fig:combined_modes}\textit{(q)}) is poorly converged. Rotor I displays localized vorticity (figure~\ref{fig:combined_modes}\textit{(r)}), whereas Rotors~II and IV yield the most prominent, well-converged harmonic structures in this band (figure \ref{fig:combined_modes}\textit{(s, t)}), marked by the largest spectral gaps.

While the frequency bands $f_{\text{bp}}$, $f_{\text{rot}}$, and $f_{\text{harm}}$ relate directly to fundamental aerodynamic and forcing frequencies, the highly energetic peak at $f_{\star}$ (dashed line, figure~\ref{fig:combined_spect}) as well as the other peaks in the band $f_{\upDelta}$ elude immediate interpretation. Occurring at frequencies that might initially be mistaken for subharmonics (i.e. $\frac{1}{2} f_{\text{rot}}$), it is straightforward to verify that these peaks are not rational multiples of either $f_{\text{rot}}$ or $f_{\text{bp}}$. Instead, the following relationship holds true for each rotor:

\begin{eqnarray}
    f_{\upDelta}^{i} = f_{\text{rot}}^{i} - f_{\star}, \quad i \in \{ \text{I}, \text{II}, \text{III}, \text{IV} \} \label{eq:delta}
\end{eqnarray}
Inspecting the mode shapes in this band reveals a striking structural difference: the mode at $f_{\star}$, shown in figure \ref{fig:combined_modes}\textit{(b)}, is supported across all four quadrants, whereas the modes in figure \ref{fig:combined_modes}\textit{(a, c--e)} clearly emphasize a single rotor each. The ordering in which the modes for each rotor appear with respect to frequency is identical to that at $f_{\text{rot}}$ and $f_{\text{bp}}$, which is consistent with the relationship established in equation \ref{eq:delta}. The modes in figure \ref{fig:combined_modes}\textit{(a, c--e)} do not consist of easily interpretable structures, such as wingtip vortices or helical vortices, implying the presence of a nonlinear aero- or vortex-dynamic mechanism. The effects of peak splitting and spectral broadening previously observed become obfuscated by the relatively large bin width and are not visible in the peaks (a, c--e).

Because the global analysis demonstrates that the vast majority of modes are localized to individual rotors, we isolate rotor-specific energy contributions by decomposing the domain into four quadrants via weighting matrices, analogous to Section~\ref{sec:frequency-time}. The SPOD modes computed in this manner represent the flow field in a single quadrant optimally in terms of variance, while also including any correlated structures in other regions of the computational domain. 

\begin{figure}
  \centerline{\includegraphics[scale = 1.0]{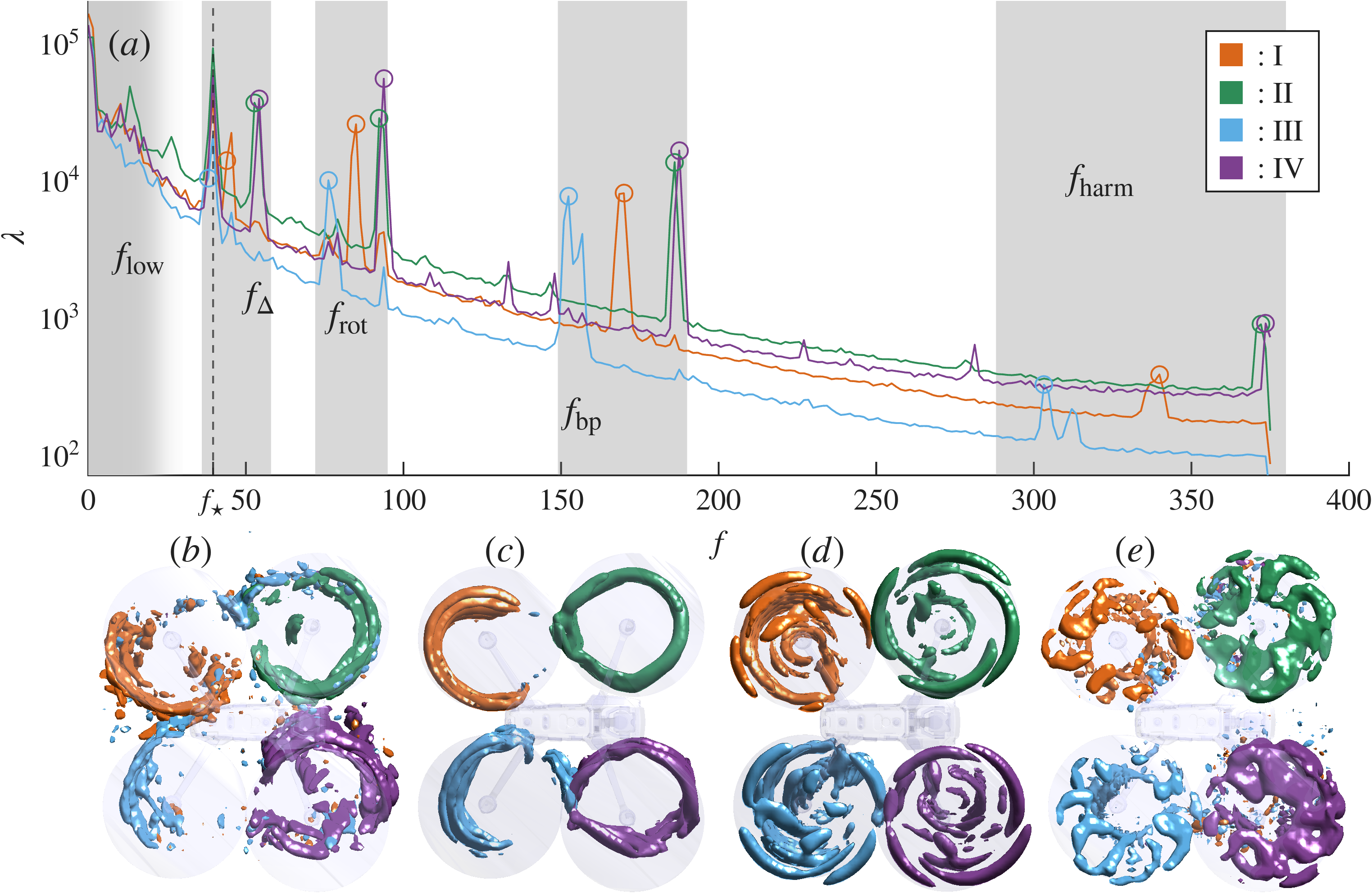}}
  \caption{\textit{(a)}: SPOD eigenvalue spectrum obtained using weighting matrices to emphasize the contributions of the four different rotors. The back rotors of the drone are denoted I (orange) and III (blue); the front rotors are II (green) and IV (purple), as shown in the schematic in figure \ref{fig:scalograms}\textit{(d)}. For better legibility, only the leading eigenvalues are plotted for each spectrum. \textit{(b--e)}: superpositions of Q-criterion isosurfaces of the modes marked with dots in the frequency groupings labeled $f_{\upDelta}$, $f_{\text{rot}}$, $f_{\text{bp}}$ and $f_{\text{harm}}$, from left to right. The exact isocontours were chosen on an individual basis to increase interpretability.}
\label{fig:leading_eig_divided}
\end{figure}

The resulting quadrant-wise spectra (figure \ref{fig:leading_eig_divided}\textit{(a)}) reveal a fairly clear separation of the wake dynamics, despite the slow drift of the drone and the expansion of the wakes across quadrant boundaries. Almost every major spectral peak is driven almost exclusively by a single rotor, with the notable exception of $f_{\star}$ and some low frequencies ($f_{\text{low}}$). This is further confirmed by superposing the quadrant-weighted Q-criterion isosurfaces (figure \ref{fig:leading_eig_divided}\textit{(b--e)}), which successfully reconstruct the coaxial tip vortices ($f_{\text{bp}}$), helical vortices ($f_{\text{rot}}$), and nonlinear structures ($f_{\upDelta}$) observed in the global analysis. 

The rotor-wise analyses also uncover low-energy features that were obscured in the global analysis. For instance, the BPF first harmonics ($f_{\text{harm}}$, figure \ref{fig:leading_eig_divided}\textit{(e)}) emerge as prominent, distinct peaks in the rotor-wise spectra, explicitly capturing the peak splitting of Rotor~III, despite being barely visible in the global spectrum. Further, the isolated spectra confirm that Rotors II and IV operate at nearly identical frequencies; their distinct, well-defined peaks across all four frequency bands account for the individual, broader peaks observed in the global spectrum.

In stark contrast to these localized coherent structures, the peak at $f_{\star}\approx 39.6$ Hz stands out as a high-magnitude eigenvalue shared across all four quadrant-wise SPOD problems. This confirms that the SPOD mode at $f_{\star}$ has energy contributions from the flow field around each rotor. While the eigenvalues at $f_{\star}$ are among the most energetic in each individual spectrum, the absolute height of each quadrant's peak is lower than in the global spectrum. Having confirmed that the coherent structure is globally supported, we can infer that this is due to the total energy contribution being split up between the four quadrants, thus making the peak in each quadrant roughly 25\% of the global value. The modes at $f_{\star}$ are not plotted, since the four SPOD problems produce nearly identical modes, as the structure is globally correlated, preventing a separation into rotor-wise coherent structures.

\begin{figure}
\centerline{\includegraphics[scale=0.99]{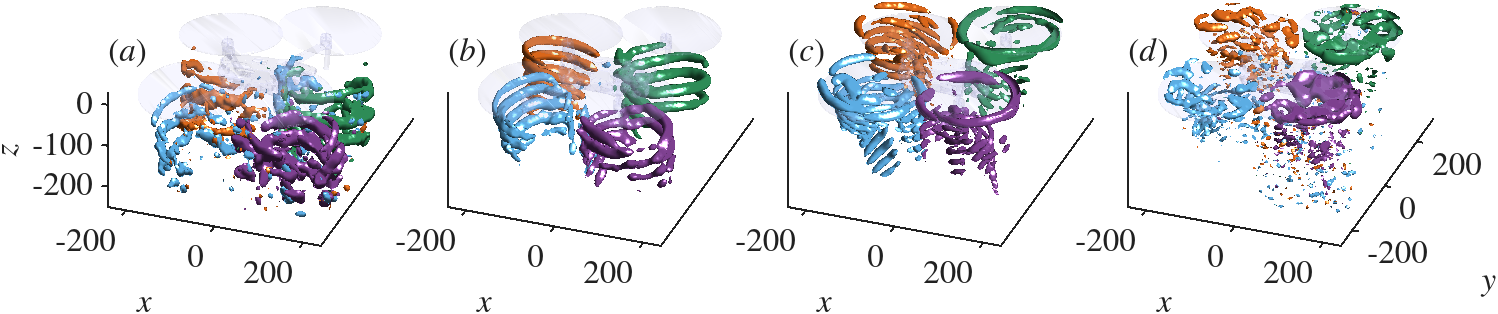}}
\caption{Q-criterion isosurfaces of most energetic SPOD modes. \textit{(a)}: $f_{\upDelta}$, \textit{(b)}: $f_{\text{rot}}$, \textit{(c)}: $f_{\text{bp}}$ and \textit{(d)}: $f_{\text{harm}}$. Q-criterion isovalues were selected on a mode-by-mode basis to increase interpretability.}
\label{fig:isometric_modes}
\end{figure}

To gain further insight into the streamwise structure of the modes, figure \ref{fig:isometric_modes}\textit{(a--d)} presents an isometric view of the same modes shown in figure \ref{fig:leading_eig_divided}\textit{(b--e)}. Examining the modes associated with $f_{\text{bp}}$ in figure \ref{fig:isometric_modes}\textit{(c)}, the Q-criterion isosurfaces reveal two coaxial systems of vortex rings per rotor. The outer ring corresponds to the well-known wingtip vortex, while the inner ring is located near the blade root, consistent with root-vortex shedding. The root vortices persist significantly farther downstream than their tip-vortex counterparts. The modes associated with the rotor frequencies $f_{\text{rot}}$, shown in figure \ref{fig:isometric_modes}\textit{(b)}, have dominant features with approximately twice the streamwise wavelength of the modes at $f_{\text{bp}}$. Unlike the modes at $f_{\text{bp}}$ and $f_{\text{harm}}$, which are directly attached to the rotor plane, the helical vortices are separated from the rotors in the streamwise direction and remain coherent for multiple revolutions. Figure \ref{fig:isometric_modes}\textit{(a)} shows the modes belonging to the $f_{\upDelta}$ frequency band, excluding the mode at $f_{\star}$. They consist of vortices with a lower spatial wavenumber compared to the modes at $f_{\text{rot}}$. The mode primarily supported around Rotor~III is not as well-converged and contains small flow structures elsewhere in the domain. Since only a single frequency bin separates this mode from $f_{\star}$, this can likely be attributed to spectral leakage. Finally, the harmonic modes in figure \ref{fig:isometric_modes}\textit{(d)}, appear less converged than those at the rotor and BPF, and are characterized primarily by vortex tube structures concentrated near the rotor planes.

\subsection{Pairing and Synchronization}
\label{sec:pairing}

\begin{figure}
\centerline{\includegraphics[width=\linewidth]{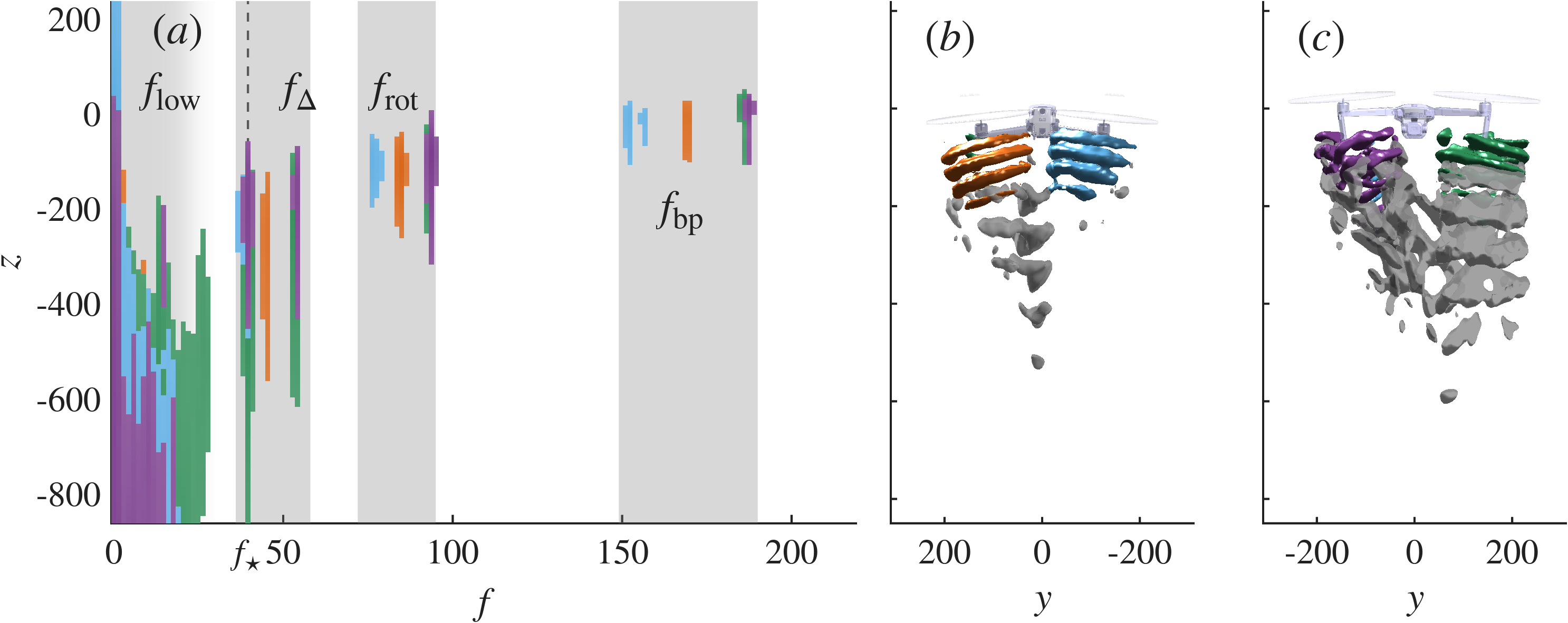}}
\caption{\textit{(a)}: Power spectrum spatially integrated across $x-$ and $y-$ coordinates at different streamwise locations. Spectral estimation parameters used are identical to those used in SPOD runs. \textit{(b)}: half-plane view of the superposed absolute magnitude isocontours of SPOD basis functions of the rotor frequencies $f_{\text{rot}}^{\text{I,III}}$ (colored opaque) and $f_{\star}$ (dark translucent). \textit{(c)}: superposed absolute magnitude isocontours of SPOD basis functions of the rotor frequencies $f_{\text{rot}}^{\text{II,IV}}$ (colored opaque) and $f_{\star}$ (dark translucent).}
\label{fig:slice_spod}
\end{figure}

The energetic significance of distinct frequency bands varies considerably with downstream distance. Figure~\ref{fig:slice_spod}\textit{(a)} maps this spatial dependency, presenting the spatially integrated power spectral density (PSD) for all four rotors along the streamwise coordinate $z$, truncated below 10\% of the maximum peak. Near the rotor planes ($z \approx 0$), the blade-passing frequencies $f_{\text{bp}}^{\text{I}-\text{IV}}$ dominate the energy spectra, reflecting the initial shedding of tip and root vortices. By $z \approx -100$, these $f_{\text{bp}}$ peaks rapidly decay, giving way to a concurrent rise in the rotor shaft frequencies $f_{\text{rot}}^{\text{I}-\text{IV}}$. Further downstream ($z \approx -200$), the $f_{\upDelta}$ frequency band becomes energetically dominant, before the low-frequency band $f_{\text{low}}$ ultimately dominates the far-field wake at $z \leq -600$.

The unstable helical vortex systems in rotor wakes often undergo a series of merging events before breaking down into turbulence \citep{leweke_long-_2014, felli_mechanisms_2011}. Interpreting the frequency band $f_{\text{bp}}$ as the fundamental frequency, subharmonic tip-vortex pairing naturally yields the $f_{\text{rot}}$ band at exactly half the frequency. Downstream, the power spectral density transitions to the frequency band $f_{\upDelta}$. Figure~\ref{fig:slice_spod}\textit{(b, c)} shows modal velocity magnitude isocontours for the frequencies $f_{\text{rot}}^{\text{I}-\text{IV}}$, colored by rotor, as well as $f_{\star}$, the dominant peak of the $f_{\upDelta}$ band, in gray, across the $x < 0$ and $x > 0$ half-planes, respectively. The $f_{\star}$ mode concentrates heavily along the longitudinal plane at $y \approx 0$ that separates the left (I, II) and right (III, IV) rotor pairs, where individual wakes first begin to interact. Whereas the frequencies $f_{\text{bp}}^{\text{I}-\text{IV}}$ and $f_{\text{rot}}^{\text{I}-\text{IV}}$ are associated with rotors, $f_{\upDelta}$ contains the dominant global synchronization mode $f_{\star}$, as well as rotor-specific modes. We will explain $f_{\star}$ as an intermittent, distinctly non-subharmonic synchronization and pairing mode existing at a frequency incommensurate with any fundamental aerodynamic frequencies of the system.

To investigate spatio-temporal vortex merging dynamics, we establish a conditional averaging framework. While power spectral analysis quantifies frequency content, it inherently lacks phase information. Conditional averaging provides qualitative insight into the coupling of coherent structures across frequency bands. We first demonstrate this methodology on the frequency bands $f_{\text{bp}}$ and $f_{\text{rot}}$, where tip vortices merge into helical vortices, before applying the identical pipeline to the synchronized pairing occurring at the incommensurate frequency $f_{\star}$.

Evaluation planes defined via a unit normal vector $\mathbf{n}$ approximately track the rotor vortex centerlines. We band-pass filter the out-of-plane vorticity, $\omega^{\perp} = \boldsymbol{\omega}^{\top} \cdot \mathbf{n}$, retaining only the frequency bands of interest, yielding a filtered signal $\tilde{\omega}^{\perp}$ that isolates any interactions between these frequencies. Evaluating $\tilde{\omega}^{\perp}$ along a spatial curve tracking the vortex cores generates $x-t$ diagrams, where merging events manifest as characteristic pitchfork signatures \citep{sherry_interaction_2013, nemes_mutual_2015}. Extracting peaks from $\tilde{\omega}^{\perp}$ at a downstream coordinate $s_0$ dominated by the lower target frequency yields the timestamps $\Omega_{t}$, which define an ensemble of merging events that can then be averaged:

\begin{align}
    \Omega_{t} &= \{t_{k}  \vert \;  \text{peaks}(\tilde{\omega}^{\perp}(s = s_{0},t_{k}))\} \label{eq:sampling} \\
    \mathbf{q}_{\text{m}}(\mathbf{x}, \tau) &= \frac{1}{N_{\Omega_{t}}}\sum_{t_k \in \Omega_{t}}\mathbf{q}(\mathbf{x},t_{k}+ \tau) \hspace{0.2in} \tau \in [-50\upDelta t, 10 \upDelta t] \label{eq:averaging}
\end{align}
Using these timestamps $\Omega_{t}$ to average both the filtered ($\tilde{\omega}^{\perp}$) and raw ($\omega^{\perp}$) out-of-plane vorticity confirms that the identified pairing mechanism accurately reflects the underlying physical dynamics.

\begin{figure}
\centerline{\includegraphics[width=\linewidth]{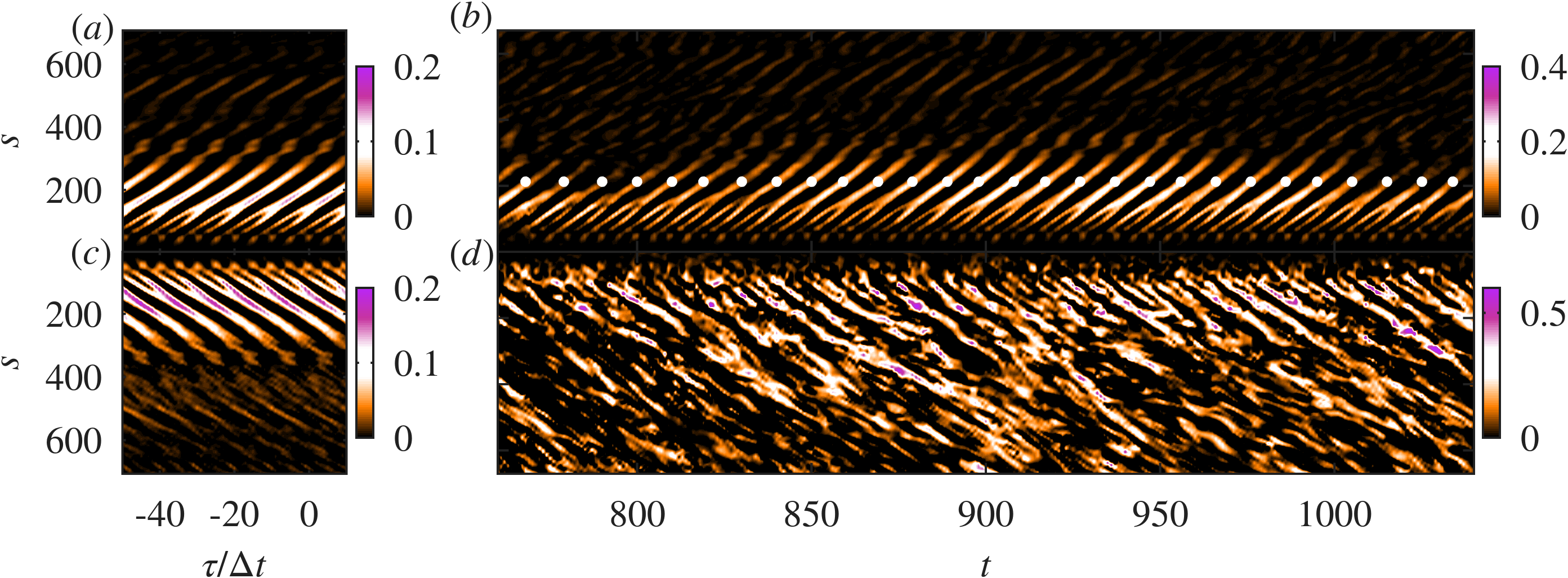}}
\caption{\textit{(a, c)}: conditional average of filtered \textit{(a)} and original data \textit{(c)}. \textit{(b)}:  $x-t$ plot of $ \tilde{\omega}^{\perp} = [0.0932 \quad 0.9763 \quad 0.1953] \cdot \boldsymbol{\tilde{\omega}}$ along the wake edge of Rotor III. The dots indicate the location of some of the maxima of $\omega^{\perp}$ used for the averaging. \textit{(d)}: $x-t$ plot of $ \omega^{\perp} = [0.0932 \quad 0.9763 \quad 0.1953] \cdot \boldsymbol{\omega}$ along the wake edge of Rotor III. For ease of interpretability, only positive values of the plotted quantities are shown.}
\label{fig:average_bp}
\end{figure}

We first apply this pipeline to the $f_{\text{bp}}$ and $f_{\text{rot}}$ frequency bands to identify the merging of consecutive tip vortices. The regular merging process easily seen in the filtered $x-t$ diagram (figure~\ref{fig:average_bp}\textit{(a)}) completes by $s = 200$, where the signal transitions fully to $f_{\text{rot}}$. In contrast, the unfiltered $x-t$ diagram (figure~\ref{fig:average_bp}\textit{(c)}) is heavily obscured by noise. Conditional averaging of an ensemble detected at $s_0 = 200$ recovers the characteristic pitchfork merging signature in both filtered (figure~\ref{fig:average_bp}\textit{(b)}) and unfiltered out-of-plane vorticity data (figure~\ref{fig:average_bp}\textit{(d)}).

\begin{figure}
\centerline{\includegraphics[width=\linewidth]{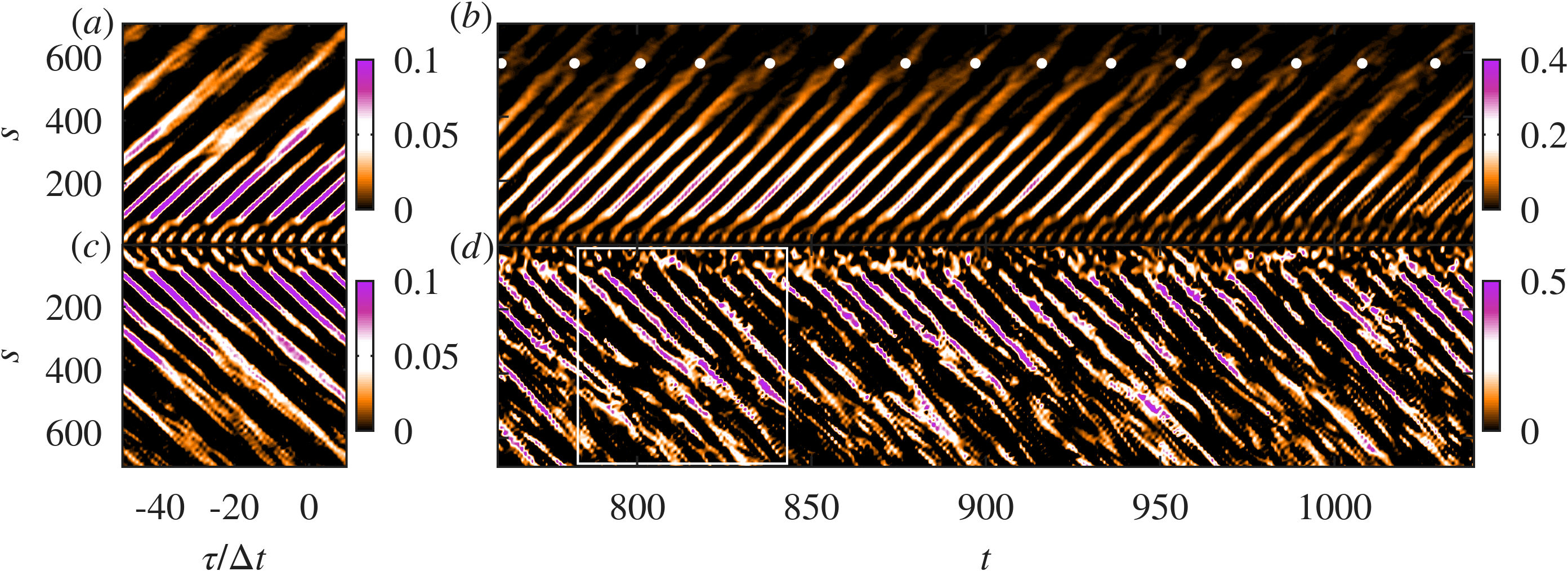}}
\caption{\textit{(a, c)}: conditional average of filtered \textit{(a)} and original data \textit{(c)}. \textit{(b)}: $x-t$ plot of the out-of-plane vorticity $ \tilde{\omega}^{\perp} = \cdot [-0.095 \quad 0.995 \quad 0]\cdot\boldsymbol{\tilde{\omega}}$ at a curve tracking the edge of the wake of Rotor II closest to Rotor I. The dots indicate the location of some of the maxima of $\omega^{\perp}$ used for the averaging. \textit{(d)}: $x-t$ plot of the out-of-plane vorticity $ {\omega}^{\perp} =  [-0.095 \quad 0.995 \quad 0]\cdot\boldsymbol{{\omega}}$. The white frame in indicates the timesteps shown in figure \ref{fig:instance}\textit{(i--p)}. For ease of interpretability, only positive values of the plotted quantities are shown.}
\label{fig:average}
\end{figure}

Having demonstrated how the averaging routine recovers the expected merging of tip vortices, we now shift our attention to the synchronized pairing between the $f_{\text{rot}}$ frequency band and the single, incommensurate frequency $f_{\star}$. In the filtered $x-t$ diagram (figure~\ref{fig:average}\textit{(a)}), the dominant frequency of the out-of-plane vorticity $\omega^{\perp}$ transitions from $f_{\text{rot}}$ to $f_{\star}$ at $s \approx 400$.

\begin{figure}
\centerline{\includegraphics{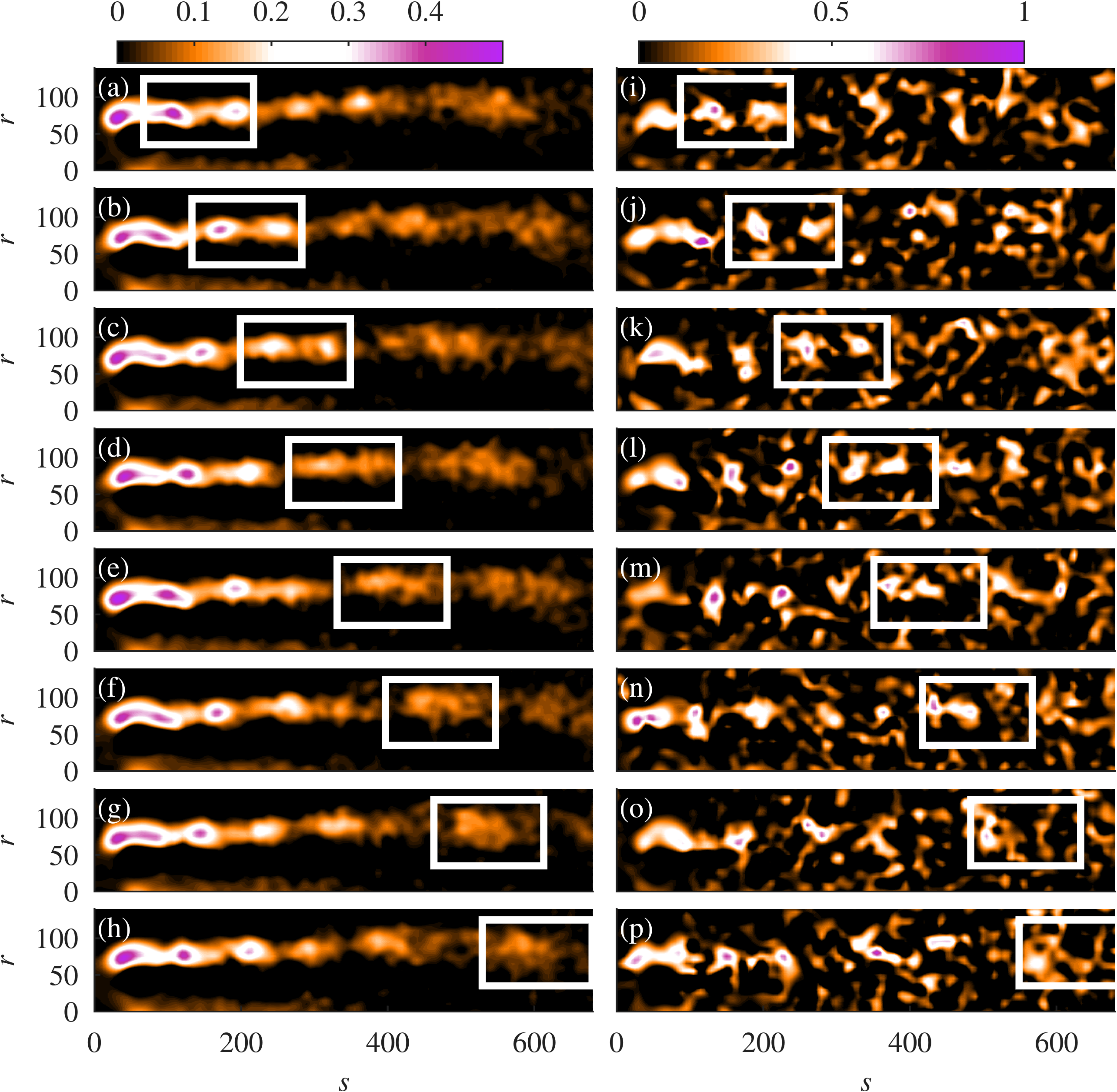}}
\caption{Visualization of the conditional average \textit{(a--h)} and a single vortex merging event \textit{(i--p)} in the wake of rotors I and II, visualized with the out-of-plane vorticity $ \omega^{\perp} = [-0.095 \quad 0.995 \quad 0] \cdot \boldsymbol{\omega}$. The local coordinates of the plane are denoted $r$ and $s$. The box is moving at a constant speed approximately equal to the phase velocity of the vortices. We note that the colormap of \textit{(a-h)} is saturated at half the magnitude, and only positive values of the quantities are shown, both solely for ease of interpretability of the plots.}
\label{fig:instance}
\end{figure}

Conditional averages of events detected at $s_0 \approx 550$ for both filtered (figure~\ref{fig:average}\textit{(b)}) and unfiltered data (figure~\ref{fig:average}\textit{(d)}) reveal a distinct merging event. To further aid interpretability, figure~\ref{fig:instance}\textit{(a--h)} depicts a two-dimensional view of the ensemble-averaged dynamics: two well-defined vortices migrate downstream before coalescing into a single, weaker, yet coherent core. In contrast, the individual ensemble member shown in figure~\ref{fig:instance}\textit{(i--p)} displays highly irregular, fluctuating vortex structures where the merging process is nearly impossible to observe directly. This stark contrast underscores the utility of conditional averaging.

The unfiltered $x-t$ diagram (figure~\ref{fig:average}\textit{(c)}) demonstrates that this pairing is intermittent; individual higher-frequency vortices occasionally exit the evaluation plane, potentially accompanied by an acceleration of another high-frequency vortex. Because $f_{\star}$ is strictly incommensurate with all four rotor frequencies, merging cannot occur at exactly subharmonic frequencies, ruling out period-doubling. We verified the essential role of $f_{\star}$ through an exclusion test: not including $f_{\star}$ in the band-pass filter completely eliminated all traces of merging from the conditional average. For brevity, we present data from a representative region at the wake interface between rotors I and II, although similar behavior occurs across the rotor wakes.

\section{Discussion}
\label{sec:discussion}

The modal decomposition of the full-scale quadcopter wake recovers fundamental aerodynamic structures documented in the literature, providing a baseline for the novel multi-rotor dynamics. Specifically, coaxial vortices originating from blade tips and roots at $f_{\text{bp}}$, together with helical vortices offset from the rotor planes oscillating at $f_{\text{rot}}$, align closely with modal analyses of wind turbines and propellers \citep{debnath_towards_2017, magionesi_modal_2018}. Unlike ideal axisymmetric configurations, the vortex structures here exhibit spatial asymmetry caused by rotor plane tilt and mean cross-flow, mirroring oblique propeller wake distortions \citep{felli_propeller_2018}. Additionally, $x-t$ diagrams and conditional averaging in Section~\ref{sec:pairing} demonstrate subharmonic pairing of tip vortex pairs at $f_{\text{bp}}$ yielding the helical vortices at $f_{\text{rot}}$, matching established single-rotor tip-vortex interactions \citep{nemes_mutual_2015, magionesi_modal_2018}.

The primary contribution of this work is the characterization of a non-subharmonic vortex merging process occurring between two vortices from the rotor frequency band $f_{\text{rot}}$ yielding vortices at the isolated, incommensurate frequency $f_\star \approx 39.6\text{ Hz}$. Because $f_\star$ is incommensurate with the fundamental aerodynamic frequencies of individual rotors, this interaction cannot stem from a conventional single-rotor subharmonic. This non-subharmonic merging departs significantly from classical symmetric vortex pairing. Rather than the symmetric acceleration and deceleration typical of subharmonic pairing, the interaction exhibits irregularly varying degrees of asymmetry: one vortex core experiences localized velocity fluctuations and core widening, while the adjacent core rapidly loses intensity. Although qualitative visual evidence of asymmetric pairing has been noted in single-rotor wakes \citep{sherry_interaction_2013, nemes_mutual_2015}, those studies did not characterize potential spectral signatures. Here, despite the asymmetry and intermittency, conditional averaging robustly demonstrates that the coherent structure at $f_\star$ is directly tied to the coalescence of two helical vortices within the $f_{\text{rot}}$ band.

The origin of the frequency $f_\star$ eludes explanation via established linear or triadic mechanisms. For axisymmetric jet flows, Floquet analysis was successfully employed to predict the onset and parametric boundaries of subharmonic growth associated with vortex merging \citep{shaabani-ardali_vortex_2019}. For subharmonic vortex merging in shear layers, three-wave resonances were empirically detected from experimental data using the bispectrum \citep{hajj_subharmonic_1992}, describing the nonlinear interactions characteristic of a saturated subharmonic resonance. The coexistence of four independent fundamental aerodynamic frequencies precludes a Floquet analysis of this flow due to the lack of a periodic base flow. The interaction between $f_{\star}$ and the band $f_{\text{rot}}$ is also not triadic, rendering bispectral methods \citep{schmidt_bispectral_2020, yeung_revealing_2024} not applicable.

Instead, the evidence points to a novel synchronization mechanism. In dynamical systems theory, synchronization is typically observed as either a consequence of external forcing, or as a result of mutual (two-way) coupling of multiple self-sustaining oscillators. Viewing the individual rotor wakes as mutually coupled systems provides a plausible interpretation of $f_\star$ as the result of a synchronization process \citep{pikovsky2003universal, balanov2008synchronization}. In this framework, inter-wake coupling likely produces sporadic episodes of phase alignment and subsequent phase slip. This intermittent coupling prevents the establishment of a phase-locked subharmonic state for any of the rotors, leading to the irregular, asymmetric merging process. Since the dataset only includes post-transient dynamics, we observe the frequency $f_\star$ as a steady-state manifestation of the synchronization process \citep{pikovsky2003universal, balanov2008synchronization}. Further, it appears that the coupling does not affect the rotor frequency band $f_{\text{rot}}$, whose frequencies are constant and experience no detuning in the streamwise direction. In the absence of parametric data, it was not possible to rigorously infer the parameters of a system of coupled oscillators to recover the frequency-domain behavior of the synchronization as in \citet{yang_computational_2025}. 

Extracting the physical insights gained in this work required navigating the inherent imperfections of experimental data. Unlike numerical simulations or heavily constrained experiments, the dataset features random, low-frequency vehicle drifts, as well as unknown control action, due to the controller maintaining hover flight. As discussed in Section~\ref{sec:frequency-time}, continuous control feedback induces independent, time-varying motor speed fluctuations across the rotors, giving rise to spectral broadening and peak splitting within the primary frequency bands. Furthermore, low-frequency vehicle drift manifests as elevated energy in the zeroth frequency bin; we confirmed that subtracting a block-wise mean when computing the SPOD significantly decreases the energy in this bin, but does not affect the other eigenvalues, and therefore also does not alter any of the relevant dynamics. Finally, the analysis was constrained by the spatio-temporal resolution at hand. Because active control inputs continuously alter the motor speeds, phase-locking the sampling to a single rotor is impossible in practice. Overcoming this lack of synchronization to approximate a co-rotating frame would require a very high temporal sampling rate. Resolving fine-scale sub-vortical mechanics, such as elliptical instabilities or vortex sheet stretching, would require significantly higher particle seeding densities \citep{wolf_experimental_2019}.

Instead of weakening the main conjecture of this work, experimental complexities ultimately reinforce the physical robustness of the underlying fluid dynamics. Crucially, $f_\star$ corresponds to the highest energy peak across the entire spectrum, establishing that the synchronized vortex merging is not only detectable, but represents a highly energetic coherent feature that dominates large portions of the wake. The spontaneous emergence of this peak within a well-resolved yet untuned experimental configuration, along with its persistence despite active control inputs, demonstrates that synchronized vortex merging is a resilient and fundamental physical process. 

\section{Conclusion and Outlook}
\label{sec:conclusion}

In this work, modal decomposition techniques are applied to the flow field around a hovering, full-scale quadcopter, obtained from data-assimilated high-density LPT. Characterized by independent, time-varying motor speeds, the observed flow exhibits multi-tonal dynamics distinct from canonical fixed-RPM single-rotor configurations.

Our analysis using SPOD successfully recovers the fundamental aerodynamic features of the drone, including the coaxial tip and root vortices, as well as the vortex merging connecting the $f_{\text{bp}}$ and $f_{\text{rot}}$ frequency bands. Further, we identified and characterized a non-subharmonic vortex merging process occurring at an incommensurate frequency $f_\star$. We demonstrate that the coherent structure associated with this frequency plays a crucial role in an intermittent, often asymmetric, synchronized merging of the four rotor wakes, fundamentally driven by nonlinear coupling. Conditional averaging extracts an interpretable merging event from the dataset. Future work should proceed along three primary avenues:

First, there is a need to establish a canonical flow capable of reproducing the synchronized merging observed in this dataset, either in simulations or experiments. The literature on fluid-dynamic synchronization remains surprisingly sparse; existing studies predominantly focus on externally forced flows \citep{li_phase_2013, herrmann_modeling_2020, provansal_benard-von_1987}, reactive systems \citep{guan_synchronization_2022, bonciolini_low_2021}, or aeroacoustics \citep{pagliaroli_rotor_2025, zarri_effects_2025}, rather than the mutual (two-way) coupling of large-scale flow structures observed in this work. A promising starting point could be a simplified configuration consisting of two mutually interacting helical vortices. This flow would include the two key phenomena of vortex merging and mutual coupling, allowing for a systematic investigation into the parametric boundaries of the synchronization process. These boundaries likely depend on the spatial separation distance (influencing coupling strength) and the frequency detuning between rotors. Furthermore, absolute tip speed ratios dictate the longevity of individual tip vortices prior to merging \citep{felli_mechanisms_2011}. Exploring and mapping this parameter space would help characterize the dynamics of multi-tonal vortex flows.

Second, future work should focus on understanding the fundamental vortex dynamical mechanisms underlying the mutual interactions. Mutual inductance mechanisms have been employed successfully to explain both leapfrogging in coaxial helical vortices and long-wave instabilities. It is plausible that an inductance-based mechanism is responsible for the mutual coupling and subsequent synchronization, and persists in bounded regions of the parameter space, providing a kinematic foundation for the vortex interactions.

Finally, the acquisition of rich parametric data would enable the formulation of a low-order mathematical model, treating the interacting wakes as a network of coupled nonlinear oscillators \citep{herrmann_modeling_2020, yang_computational_2025}, inferring both functional form and model parameters from data. In contrast to infinite-dimensional systems such as the Navier--Stokes equations, established methods from the dynamical systems community exist for rigorous investigations into ODEs. Similar to \cite{li_phase_2013}, an oscillator model could be used to perform a parametric analysis of the phase space and a classification of potential bifurcation routes to synchronization, providing generalizable qualitative insights into the emergence of synchronization in multi-tonal vortex flows.

By leveraging a tailored approach to discover evidence of a previously unexplored physical mechanism, this work highlights the capacity of spectral modal decomposition to extract physically interpretable structures from full-scale experimental data, in the face of typical uncertainties.

\section*{Acknowledgements}
The authors gratefully acknowledge support from the Army Research Office under Grant No. W911NF-25-1-0222, with Dr. Kenneth Granlund serving as Program Manager. We also thank Dr. Granlund for facilitating our connection with our collaborators at DLR and helping initiate this fruitful collaboration.

\section*{Declaration of Interests}
The authors report no conflict of interest.

\bibliographystyle{jfm}
\bibliography{paper_final}

\end{document}